\documentclass[11pt, a4paper]{article}
\usepackage{jcappub}

\usepackage{verbatim}
\usepackage{physics}

\usepackage{bm}

\allowdisplaybreaks[4]

\title{Window function dependence of the novel mass function of primordial black holes}
\author[a, b]{Koki Tokeshi,}
\author[b]{Keisuke Inomata,}
\author[a, b, c]{and Jun'ichi Yokoyama}
\affiliation[a]{
Department of Physics, 
Graduate School of Science, 
The University of Tokyo, \\
7-3-1 Hongo, Bunkyo, Tokyo, 113-0033, Japan
}
\affiliation[b]{
Research Center for the Early Universe (RESCEU), 
Graduate School of Science, \\
The University of Tokyo, 
7-3-1 Hongo, Bunkyo, Tokyo, 113-0033, Japan
}
\affiliation[c]{
Kavli Institute for the Physics and Mathematics of the Universe (Kavli IPMU), \\
The University of Tokyo, 
5-1-5 Kashiwanoha, Kashiwa, Chiba, 277-8583, Japan}
\emailAdd{tokeshi@resceu.s.u-tokyo.ac.jp}
\emailAdd{inomata@resceu.s.u-tokyo.ac.jp}
\emailAdd{yokoyama@resceu.s.u-tokyo.ac.jp}
\abstract{
We investigate the ambiguity of the novel mass function of primordial
black holes, which has succeeded in identifying the black hole mass in a
given configuration of fluctuations,
due to the choice of window function of smoothed density
fluctuations.
We find that while the window function dependence of the exponential factor
in the novel mass function is the same as the one in the conventional mass
function around the top-hat scale, the dependences are different on 
other scales,
which leads to the narrower mass function in the novel formulation for 
some window
functions.}
\arxivnumber{XXXX.XXXX}

\begin{document}

\maketitle

\section{Introduction}

Primordial black holes (PBHs) are produced in the early Universe when the density fluctuations larger than the threshold enter the Hubble horizon~\cite{1967SvA....10..602Z,Hawking:1971ei,74CH}.
The formation of PBHs and its cosmological implications have been discussed from a variety of viewpoints for several 
decades~\cite{75Carr,1977PAZh....3..208Z,1978SvA....22..129N,Kohri:1999ex,Carr:2009jm}.
PBHs are candidates for a portion of the dark matter (DM) and some heavy BHs recently observed by the LIGO-Virgo collaboration~\cite{96KL, 12DE, 10LB, 16CKS, Bird:2016dcv,Clesse:2016vqa,Sasaki:2016jop,Inomata:2016rbd,17IKMTY,Inomata:2017vxo,19Carr}.
In addition, the abundance of PBHs can give us some clues of small-scale density fluctuations that are related to properties of inflation models but not accessible by cosmic microwave background (CMB) or the large-scale structure observations~\cite{Carr:1993aq,Carr:1994ar,Josan:2009qn,Cole:2017gle,Carr:2017edp,Dalianis:2018ymb,Sato-Polito:2019hws,Kalaja:2019uju}. (See also refs.~\cite{18SSTY,Carr:2020gox} for recent reviews on PBHs.)

The relation between the power spectrum of curvature perturbations, related to the inflation models, and the PBH mass function has been studied in refs.~\cite{,Kim:1996hr,04GLMS,14YBS,Yoo:2018kvb,19GM,Atal:2018neu} with the use of the Press--Schechter formalism~\cite{press1974formation} or the peaks theory~\cite{86BBKS}.
However, as pointed out in ref.~\cite{20SY}, the mass functions so far have some issues at the conceptual level, which stem from the fact that the conventional PBH formation criteria are not enough for straightforwardly giving the mass function, defined as a differential quantity. 
Recently the authors in ref.~\cite{20SY} have deduced that a PBH is produced around a certain peak at the mass scale where smoothed density perturbation around it takes the maximum value as the smoothing scale $R$ and corresponding mass scale are changed.
From this observation they have proposed a new additional criterion for PBH formation to avoid the issues, 
which is to require the smoothed density fluctuation to be maximal with the change of $R$.
With this additional criterion, they have succeeded in defining a novel mass function.
Also, the same criterion is proposed in ref.~\cite{Germani:2019zez}.
Although the prefactor of the mass function in ref.~\cite{Germani:2019zez} is different from the one in ref.~\cite{20SY}, which comes from the difference of other imposed conditions, the exponential factors in both mass functions are similar to each other.

Since PBH formation is a phenomenon occurring on the horizon scale at the formation time, we expect that the PBH formation should not depend on the perturbations on the deeply subhorizon.
In other words, we expect that the contribution of the subhorizon perturbations to the PBH formation must be suppressed.
To take into account this suppression, the window (or filter/smoothing) function has been introduced in many references.\footnote{
In ref.~\cite{Yoo:2018kvb}, the authors do not use the window function and, instead relate a PBH mass with the height and the curvature of the peak. However, the analysis cannot be applied to broad or multiple-peak power spectra of curvature perturbations~\cite{Yoo:2018kvb}.
}
However, there is no consensus on the choice of the window function so far.
The window function dependence of the conventional mass function has been studied on in refs.~\cite{18AIK,20Young}, which 
focus on the window function dependences of the smoothed density contrast~\cite{18AIK,20Young} and the threshold~\cite{20Young}.
However, unlike the conventional mass function, the novel mass function includes in its exponential factor the density contrasts differentiated with respect to the smoothing scale, as we will see in section~\ref{sec:basic_formulas}.
Since the mass function sensitively depends on its exponential factor, it is worthwhile to revisit the window function dependence of the factor.
In this paper, we discuss the window function dependence, specifically taking the following three window functions, which are  used in the literature, namely,  the Gaussian window function and the top-hat window functions in position space and wavenumber space which
we call $x$-space top-hat and  $k$-space top-hat window functions, respectively.
Then, we report the updated window function dependence of the exponential factor.

The rest of the paper is organized as follows. 
In section~\ref{sec:basic_formulas}, we review the basic formulas needed to carry out our discussion. 
Then, we review the three kinds of window functions and the behavior of the smoothed density contrast in section~\ref{sec:review_on_window}. 
In section~\ref{sec:window_novel}, we discuss the window function dependence of the novel mass function.
Finally, we devote section~\ref{sec:conclusion} to conclusion and discussion.

% ///// ///// Section 2 ///// /////
\section{Basic formulas for PBH formation}
\label{sec:basic_formulas}

In this section, we briefly summarize the basic formulas for PBH formation, which are mainly based on the formulation recently reported in ref.~{\cite{20SY}}.
Note that we assume that PBHs are produced during the radiation-dominated era throughout this paper.

\paragraph{Density contrast}
One of the useful quantities to describe PBH formation is the density contrast on the comoving slice, which is defined as $\delta \equiv \delta \rho/\bar \rho$ with $\bar \rho$ and $\delta \rho$ being the background and the perturbation of the energy density, respectively.
We can express the density contrast in terms of the curvature perturbation $\mathcal{R}$ at linear order as\footnote{The effects of the non-linear relation between $\delta$ and $\mathcal R$ on the PBH abundance have been discussed in refs.~\cite{Kawasaki:2019mbl,DeLuca:2019qsy,Young:2019yug,Mahbub:2020row}.}
\begin{equation}
\delta (\vb*{x})
\simeq - \frac{4}{9} \frac{1}{ ( aH )^2 } \nabla^2 \mathcal{R} (\vb*{x}) \,\, ,
\label{eq:2_1}
\end{equation}
where this relation is valid on superhorizon scales and in the radiation-dominated era. 
Here, $1/(aH)$ denotes the conformal Hubble horizon with $a$ and $H$ being the scale factor and the Hubble parameter, respectively. 
Since the PBH formation is expected not to be affected by modes deep inside the horizon, 
it should be described by the density contrast that is coarse-grained through some window function $W (R, r)$ with the smoothing scale $R$ being the horizon scale $1/(aH)$ as
\begin{align}
\delta (R, \vb*{x})
&\equiv \int \dd^3 y \, W (R, |\vb*{y} - \vb*{x}|) \delta (\vb*{y}) \notag \\
&\simeq - \frac{4}{9} R^2 \int \dd^3 y \, W (R, |\vb*{y} - \vb*{x}|) \nabla^2 \mathcal{R} (\vb*{y}) \,\, .
\label{eq:2_2}
\end{align}
Hereafter, we call $\delta (R, \bm x)$ the smoothed density contrast and denote it by $\delta_R (\vb*{x})$ or $\delta_R$ for short.
The concrete expressions of the window functions will be given in section~\ref{sec:review_on_window}.

\paragraph{Power spectrum of curvature perturbation} 
The power spectrum of the curvature perturbation is defined with their Fourier modes, $\tilde{\mathcal{R}}$, as
\begin{align}
	\expval{\tilde{\mathcal{R}}(\bm k) \tilde{\mathcal{R}}(\bm k')} = (2\pi)^3  \delta_\text{D} (\bm k + \bm k') \frac{2\pi^2}{k^3}  \mathcal P_{\mathcal R}(k),
	\label{eq:p_r}
\end{align}
where the angle brackets stand for the ensemble average, $\delta_{\mathrm{D}} (x)$ is the Dirac $\delta$-function, and $k = |\bm k|$.
To make the discussion concrete, we take the top-hat power spectrum as a fiducial example throughout this paper, given by
\begin{equation}
\mathcal{P}_{\mathcal{R}} ( k )
= 
\mathcal{A} 
\Theta (k - k_l) \Theta ( - k + k_s ) 
\,\, ,
\label{eq:2_3}
\end{equation}
where $\mathcal{A}$ is the amplitude of the power spectrum,
$\Theta (x)$ is the Heaviside step function, 
and both $k_l$ and $k_s$ are the cut-off wavenumbers ($0<k_l<k_s$).\footnote{Throughout this paper, we just assume that the value of $\Theta(0)$ is finite, that is, do not assume any specific value of $\Theta(0)$, though it appears in Eqs.~(\ref{eq:4_5}) and (\ref{eq:4_6}).
This is because the final result of the exponential factor does not depend on the finite value of $\Theta(0)$ due to the divergent factors in Eqs.~(\ref{eq:4_5}) and (\ref{eq:4_6}).}

\paragraph{PBH mass function}
Here, we review the novel PBH mass function recently reported in ref.~\cite{20SY}, which imposes the following criteria for PBH formation:
\begin{equation}
\delta (R, \vb*{x}) > \delta_{\mathrm{th}} \,\, ,
\qquad
\pdv{\delta (R, \vb*{x})}{x^{\alpha}} = 0 \,\, ,
\qquad
\lambda_{\alpha} (H) < 0 \,\, ,
\label{eq:2_5}
\end{equation}
where $x^\alpha$ runs over the 3-dimensional space coordinates and $R$, that is, $x^i$ with $i=1,2,3$ denote the space coordinates and $x^4$ denotes the smoothing scale $R$.
$\lambda_{\alpha}$'s are the eigenvalues 
of the $4 \times 4$ Hesse matrix for $\delta (R, \vb*{x})$, that is, $H_{\alpha \beta} = \partial^2 \delta (R, \vb*{x}) / \partial x^{\alpha} \partial x^{\beta}$.
The first condition requires that 
PBH formation occurs if the smoothed density contrast exceeds some threshold 
$\delta_{\mathrm{th}}$. 
The second and the third conditions guarantee that, 
at the point of formation, $\delta (R, \vb*{x})$ becomes not only extremum but also maximum with respect to the 4-dimensional parameter space $x^\alpha$.
The novel point of this formulation is adding the smoothing scale $R$ to the coordinates and imposing the same condition even for $R$
so that each high density peak, which would eventually collapse to a PBH, would be counted as such at the proper mass scale.
The three conditions in eq.~(\ref{eq:2_5}) lead to the novel expression of the mass function with PBH mass $M$~\cite{20SY},
\begin{equation}
f (M)
= \frac{M}{n_\text{PBH}} 
\int \dd R \, \expval{
\abs{ \det H } 
\delta_{\mathrm{D}} ( M - m (R, \delta_R) ) \Theta ( \delta_R - \delta_{\mathrm{th}} ) 
\prod_{\alpha = 1}^4 \delta_{\mathrm{D}} (\delta_{R, \alpha}) \Theta (- \lambda_{\alpha})
}
\,\, ,
\label{eq:2_6}
\end{equation}
where $f(M)$ denotes the relative fraction of PBHs within unit logarithmic mass interval around $M$, which satisfies $\displaystyle \int \dd \ln M \, f (M) = 1$. 
The variables after commas represent derivatives with respect to them and $m$ and $n_\text{PBH}$ are the mass of a PBH related to each peak and the comoving number density of PBHs, respectively. 
Here, we neglect the effects of the critical phenomena~\cite{Niemeyer:1997mt,Niemeyer:1998ac,99SS,Yokoyama:1998xd} and assume that the PBH mass depends only on the smoothing scale for the sake of comparison with conventional approach.
Then, we can rewrite eq.~(\ref{eq:2_6}) as 
\begin{equation}
f (M)
= \frac{M}{n_\text{PBH} \, \partial m (R) / \partial R} I (M) \,\, ,
\label{eq:2_7}
\end{equation}
where
\begin{equation}
I (M) 
\equiv 
\expval{
\abs{\det H } \Theta (\delta_R - \delta_{\mathrm{th}}) 
\prod_{\alpha = 1}^4 \delta_{\mathrm{D}} (\delta_{R, \alpha}) \Theta (- \lambda_{\alpha}) 
} 
\,\, .
\label{eq:2_8}
\end{equation}
In the case where the probability distribution of $\delta_R$ is Gaussian,\footnote{The effects of the non-Gaussianity in the conventional formulation are discussed in refs.~\cite{Yokoyama:1998pt,Saito:2008em,Byrnes:2012yx,Young:2013oia,Nakama:2016kfq,Nakama:2016gzw,Nakama:2017xvq}.} 
the analytical expression of eq.~(\ref{eq:2_8}) can be obtained by rewriting the Heaviside step function and the Dirac $\delta$-functions in terms of the inverse Fourier transformations.
Doing so, we finally obtain 
\begin{align}
I (M) 
&= \frac{1}{\sqrt{\det (2 \pi L)}} 
\sqrt{ \frac{ 2 }{N_{55} } } 
\notag \\
&\quad 
\times 
\left[
\qty( A + B_{ij} N_{ij} + 3 C_{ijkl} N_{ij} N_{kl} - \frac{(B_{ij} + 6 C_{ijkl} N_{kl} ) N_{i5} N_{j5} }{N_{55}} 
+ \frac{ 3 C_{ijkl} N_{i5} N_{j5} N_{k5} N_{l5} }{N_{55}^2} 
)
\right.
\notag \\
&\qquad \quad 
\quad 
\times 
\frac{\sqrt{\pi}}{2} \mathrm{erfc} \qty( \sqrt{ \frac{N_{55}}{2} } \, \delta_{\mathrm{th}} ) 
\notag \\ 
&\qquad \quad 
+ \frac{\delta_{\mathrm{th}}}{\sqrt{2 N_{55}}} 
\qty( 
( B_{ij} + 6 C_{ijkl} N_{kl} ) N_{i5} N_{j5} 
+ \frac{ \delta_{\mathrm{th}} (3 + N_{55} \delta_{\mathrm{th}}^2 ) C_{ijkl} N_{i5} N_{j5} N_{k5} N_{l5} }{N_{55}} ) 
\notag \\
&\qquad \quad 
\quad 
\times 
\left.
\exp \qty( - \frac{N_{55}}{2} \delta_{\mathrm{th}}^2 ) 
\right] \,\, ,
\label{eq:2_9}
\end{align}
where $L$ is the symmetric and block diagonal $5 \times 5$ matrix defined by \begin{equation}
L \equiv \mqty( 
\expval{ ( \delta_{R, x} )^2 } & & & & \\
& \expval{ ( \delta_{R, y} )^2 } & & & \\
& & \expval{ ( \delta_{R, z} )^2 } & & \\
& & & \expval{ ( \delta_{R, R} )^2 } & \expval{ \delta_{R, R} \delta_R } \\
& & & \expval{ \delta_{R, R} \delta_R } & \expval{ \delta_R^2 } 
) \,\, ,
\label{eq:2_10}
\end{equation}
and $N$ is the inverse matrix of $L$. 
See appendix~\ref{app:4_dim_formula} for the derivation of eq.~(\ref{eq:2_9}) and the explicit expressions for $N$ and the other quantities, such as $A$, $B_{ij}$, and $C_{ijkl}$.
In the rest of this paper, we focus on the exponential factor in eq.~(\ref{eq:2_9}), given by 
\begin{equation}
\frac{N_{55}}{2} \delta_{\mathrm{th}}^2
= \frac{1}{2} N_{55} \expval{ \delta_R^2 } \frac{\delta_{\mathrm{th}}^2}{\expval{ \delta_R^2 }} 
= \frac{1}{2} \frac{ 
\expval{ \delta_R^2 } \expval{ ( \delta_{R, R} )^2 }
}{
\expval{ \delta_R^2 } \expval{ ( \delta_{R, R} )^2 } - \expval{ \delta_R \delta_{R, R} }^2 
} 
\frac{\delta_{\mathrm{th}}^2}{\expval{ \delta_R^2 }}
\,\, .
\label{eq:2_11}
\end{equation}
The abundance of PBHs strongly depends on this exponential factor.
Note that eq.~(\ref{eq:2_11}) has the same form as the $2$-dimensional toy model result obtained in ref.~\cite{20SY}.
Comparing eq.~(\ref{eq:2_11}) with the corresponding exponential factor of the conventional mass function, $(1/2) \, \delta_{\mathrm{th}}^2 / \expval{ \delta_R^2 }$~\cite{04GLMS}, 
we observe that the extra factor $N_{55} \expval{ \delta_R^2 }$ is additionally multiplied in the novel formulation. 
In other words, the deviation of $N_{55} \expval{ \delta_R^2 }$ from unity denotes the new effect appearing in the novel formulation.
The corresponding factors are given with the Fourier components of the window function, $\tilde W$, as 
\begin{align}
\expval{ \delta_R^2 }
&= \mathcal{A} \qty( \frac{4}{9} )^2  
\int_{k_l}^{k_s} \frac{\dd k}{k} \, 
(kR)^4 
T^2 ( \eta = R, k ) 
\tilde{W}^2 (R, k) \,\, ,
\label{eq:2_12}
\\
\expval{ ( \delta_{R, R} )^2 }
&= \mathcal{A} \qty( \frac{4}{9} )^2 
\int_{k_l}^{k_s} \frac{\dd k}{k} \, 
\qty( 
\frac{\partial}{\partial R} 
\qty[ 
(kR)^2 
T ( \eta = R, k ) 
\tilde{W} (R, k) 
] 
)^2 \,\, ,
\label{eq:2_13}
\\
\expval{ \delta_R \delta_{R, R} }
&=\mathcal{A} \qty( \frac{4}{9} )^2
\int_{k_l}^{k_s} \frac{\dd k}{k} \, 
(kR)^2 T ( \eta = R, k ) \tilde{W} (R, k) \frac{\partial}{\partial R} \left[(kR)^2 
T ( \eta = R, k ) 
\tilde{W} (R, k) \right] 
\,\, ,
\label{eq:2_14}
\end{align}
where we have used eq.~(\ref{eq:2_2}) and introduced the transfer function for the density perturbation to take into account its subhorizon evolution, which is given with the conformal time $\eta$ as~\cite{Dodelson:1282338}
\begin{align}
T (\eta, k) 
= 3 \frac{ \sin ( k \eta / \sqrt{3} ) - ( k \eta / \sqrt{3} ) \cos ( k \eta / \sqrt{3} ) }{ ( k \eta / \sqrt{3} )^3 }.
\end{align}
This transfer function plays an important role in the case of the $x$-space top-hat window function because the suppression of the subhorizon contribution by the window function itself is not so strong. 
On the other hand, the transfer function is not important for the Gaussian and the $k$-space top-hat window functions since these two window functions sufficiently suppress the subhorizon contribution by themselves~\cite{18AIK, 20Young},
and hence we set the transfer function to unity for these two cases for clarity of expression in the rest of this paper.\footnote{We have numerically checked that this prescription for the transfer function hardly affects the extra factor defined in eq.~(\ref{eq:4_1}). }

% \\\\\ \\\\\ Section 3 \\\\\ \\\\\
\section{Review on window functions and their properties}
\label{sec:review_on_window}

In this section, we review the three commonly used window functions: 
(i) the Gaussian, 
(ii) the $x$-space top-hat, 
and 
(iii) the $k$-space top-hat window function. 
Following the convention in ref.~\cite{20Young}, we define these window functions as 
\begin{enumerate}
\item[(i)]
Gaussian window function:
\begin{align}
W(R,r) &= \frac{1}{(\pi R^2)^{3/2}} \exp\left( - \frac{r^2}{R^2} \right)\,\, , \\
\tilde{W} (R, k)
&= \exp \left[ - \frac{(kR)^2}{4} \right] \,\, ,
\label{eq:3_1}
\end{align}

\item[(ii)]
$x$-space top-hat window function:
\begin{align}
W(R,r) &= \frac{3}{4 \pi R^3} \Theta (R-r)\,\, , \\
\tilde{W} (R, k)
&= 
3 \frac{ \sin (k R) - k R \cos (kR) }{ (k R)^3 } 
\,\, ,
\label{eq:3_2}
\end{align}

\item[(iii)]
$k$-space top-hat window function:
\begin{align}
W(R,r) &= 
\frac{\alpha^3}{2 \pi^2 R^3} \frac{ \sin \qty[ \alpha (r/R) ] -  \alpha (r/R) \cos \qty[ \alpha (r/R) ]}{ \qty[ \alpha (r/R) ]^3}\, \, , \\
\tilde{W} (R, k)
&= \Theta \left( \frac{\alpha}{R} - k \right) \,\, ,
\label{eq:3_3}
\end{align}
\end{enumerate}
where $\alpha = 2.744$ and the three window functions are defined such that $4 \pi r^2 W (R, r)$, which will appear in eq.~(\ref{eq:3_8}), has its peak at $r = R$~\cite{20Young}.\footnote{
Due to this peak condition for $4 \pi r^2 W (R, r)$, eqs. (\ref{eq:3_1}) and (\ref{eq:3_3}) are slightly different from the conventional window functions, used in e.g. ref.~\cite{18AIK}.
}
All the window functions above satisfy the normalization condition $\tilde{W} (R, k = 0) = 1$ in the Fourier space and the Gaussian and the $x$-space top-hat window function satisfy $\displaystyle \int \dd^3 x \, W (R, \abs{ \vb*{x} }) = 1$ in the real space. 
Substituting eqs.~(\ref{eq:3_1}), (\ref{eq:3_2}), and (\ref{eq:3_3}) into eq.~(\ref{eq:2_12}) 
and performing the variable transformation $u \equiv k R$,  
we can express $\expval{ \delta_R^2 }$ for each window function as 
\begin{enumerate}
\item[(i)]
Gaussian window function:
\begin{align}
\expval{\delta_R^2}
= \mathcal{A} \qty( \frac{4}{9} )^2 \int_\xi^{s \xi} \dd u \, u^3 e^{- u^2/2} 
= - 2 \mathcal{A} \left( \frac{4}{9} \right)^2 \Delta \Gamma_2 \,\, ,
\label{eq:3_4}
\end{align}

\item[(ii)]
$x$-space top-hat window function:
\begin{equation}
\expval{ \delta_R^2 }
= 
\mathcal{A} \qty( \frac{4}{9} )^2 \int_{\xi}^{s \xi} \dd u \, 
u^3 \qty( 3 \frac{ \sin u - u \cos u }{u^3} )^2 
\qty[ 3 \frac{ \sin ( u / \sqrt{3} ) - ( u / \sqrt{3} ) \cos ( u / \sqrt{3} ) }{ (u / \sqrt{3})^3 } ]^2 
\,\, ,
\label{eq:3_5}
\end{equation}

\item[(iii)]
$k$-space top-hat window function:
\begin{equation}
\expval{ \delta_R^2 }
= 
\begin{cases}
\displaystyle \frac{\mathcal{A}}{4} \qty( \frac{4}{9} )^2 (s^4 - 1) \xi^4 
&\quad (0 < \xi < \alpha/s)
\\
\displaystyle  \frac{\mathcal{A}}{4} \qty( \frac{4}{9} )^2 (\alpha^4 - \xi^4) 
&\quad (\alpha/s \leq \xi \leq \alpha)
\\
0
&\quad (\xi > \alpha)
\end{cases}
\,\, , 
\label{eq:3_6}
\end{equation}
\end{enumerate}
where we have introduced $\xi \equiv k_l R$ and $s \equiv k_s / k_l$, so that $k_s R = s \xi$.
Note again that we have set $T(k,\eta(=R)) = 1$ for the Gaussian and the $k$-space top-hat window functions for clarity.  Here
$\Delta \Gamma_q$ is defined as $\Delta \Gamma_q \equiv \Gamma (q, (k_s R)^2 / 2 ) - \Gamma (q, (k_l R)^2 / 2 ) = \Gamma (q, (s \xi)^2 / 2) - \Gamma (q, \xi^2 / 2)$ in terms of the incomplete gamma function, 
\begin{equation}
\Gamma (q, x) \equiv 
\int_x^{\infty} \dd t \, t^{q-1} e^{-t} \,\, .
\label{eq:3_7}
\end{equation}

To see how these window functions smooth the density contrast, 
let us observe the relation between the traditional exponential factor $\delta_{\mathrm{th}}^2 / \expval{ \delta_R^2 }$ and $\xi$. 
Here, we should notice that the threshold $\delta_{\mathrm{th}}$ also depends on the choice of the window function, as discussed in ref.~\cite{20Young}. 
Following the discussion in the reference, we assume that the threshold is given by 
\begin{equation}
\delta_{\mathrm{th}}
= \int_0^{\infty} \dd r \, 4 \pi r^2 W (R, r) \delta_\text{c} (R, r) \,\, .
\label{eq:3_8}
\end{equation}
Here we choose $\delta_\text{c} (R, r)$ as a typical density profile of the peak, which may be almost independent of the power spectrum of the perturbations~\cite{19GM,20Young}, with the numerically determined critical amplitude $A_\text{c} \simeq 1.2$~\cite{19GM}:
\begin{equation}
\delta_\text{c} (R, r) 
= A_\text{c} \frac{ \sin \qty[ \alpha ( r/R ) ] }{ \alpha ( r/R ) } \,\, .
\label{eq:3_9}
\end{equation}
The numerical results of the thresholds are listed in table~\ref{tab:3_1}~\cite{20Young}. 

\begin{figure}[h]
\begin{center}
\includegraphics[keepaspectratio, scale=0.7]{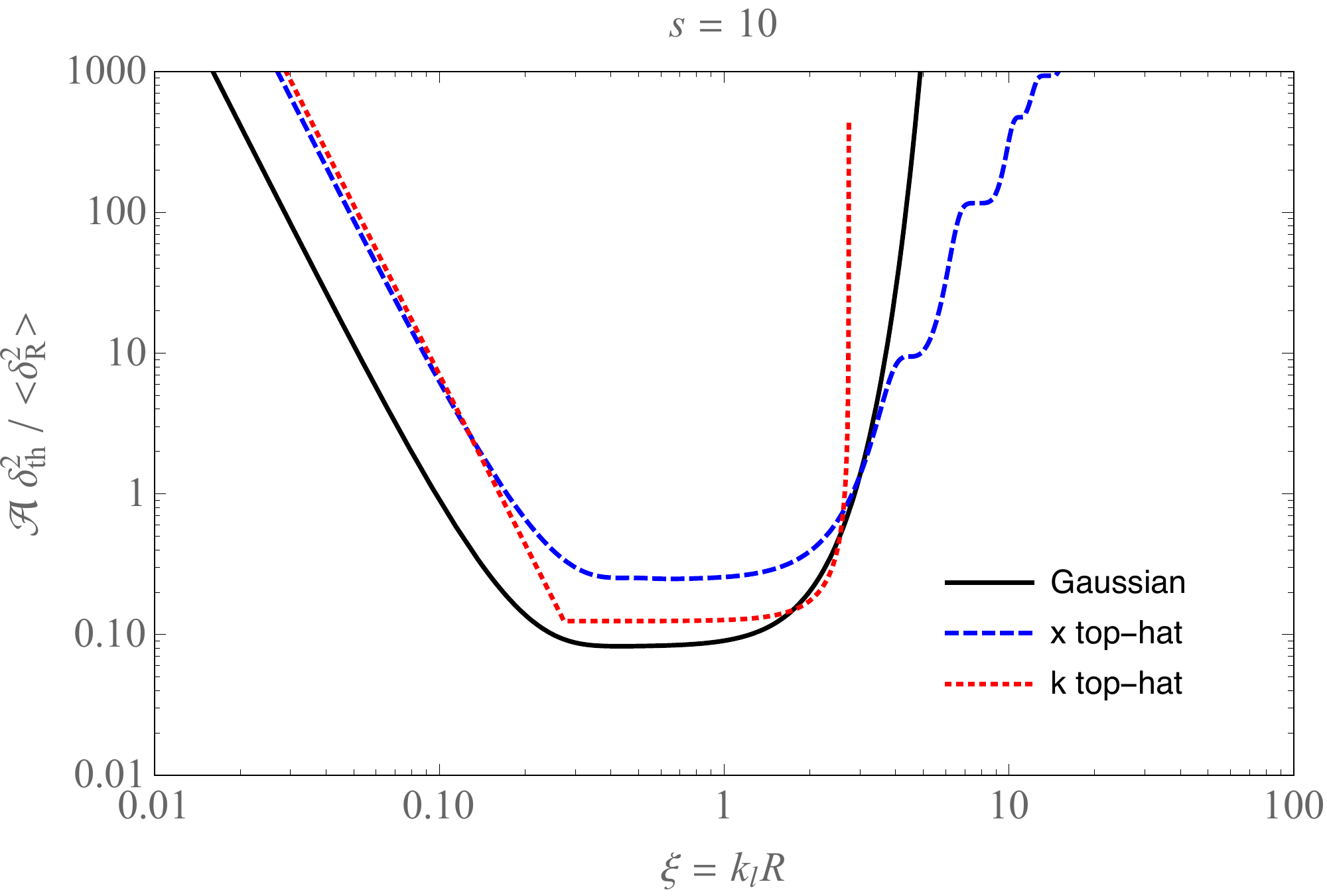}
\caption{
The behaviors of $\mathcal A\, \delta_{\mathrm{th}}^2/\expval{\delta_R^2}$ with respect to $\xi = k_l R$ in $s =10$. 
}
\label{fig:3_1}
\end{center}
\end{figure}

Figure \ref{fig:3_1} shows the behaviors of $\mathcal{A} \delta_{\mathrm{th}}^2 / \expval{ \delta_R^2 }$ at $s = 10$, 
for the three window functions.\footnote{
Note that the behavior of $\expval{ \delta_R^2 }$ is almost the same for larger $s$. 
This is also true for the results given in the next section. 
} 
It is seen that all the window functions exhibit flat behaviors in $\alpha / s \lesssim \xi \lesssim 1$. 
The smallest flat value of $\delta_{\mathrm{th}}^2 / \expval{ \delta_R^2 }$ is obtained for the Gaussian window function ($\delta_{\mathrm{th}}^2 / \expval{ \delta_R^2 } \simeq 0.082$), 
while the largest one for the $x$-space top-hat window function ($\delta_{\mathrm{th}}^2 / \expval{ \delta_R^2 } \simeq 0.24$). 
The result for the $k$-space top-hat window function is intermediate ($\delta_{\mathrm{th}}^2 / \expval{ \delta_R^2 } \simeq 0.12$) with the cut-off at $\xi = \alpha$. 

For the Gaussian window function with $\xi \gtrsim 1$, $\expval{ \delta_R^2 }$  behaves as
\begin{align}
\expval{ \delta_R^2 } \simeq \mathcal{A} \left( \frac{4}{9} \right)^2 (2 +\xi^2) e^{- \xi^2/2} 
\qquad \,\,
(\xi \gtrsim 1 \ \text{for Gaussian}) 
\,\, .
\end{align}
For the $x$-space top-hat window function with $\xi \gtrsim 1$, $\expval{ \delta_R^2 }$   can be expressed as
\begin{align}
\expval{ \delta_R^2 } 
\simeq 
&\frac{9\mathcal A}{2 \xi^5} 
\left( 
2 \xi - 4 \sin(2 \xi) - 4 \sqrt{3} \sin \left( \frac{2\xi}{\sqrt{3}} \right)
\right. %\nonumber \\
%& \qquad \ 
- ( 3 + \sqrt{3} ) \sin \qty[ \frac{2}{3} ( 3 - \sqrt{3} ) \xi ] \notag \\
& \qquad \ 
\left. 
- \, ( 3 - \sqrt{3} ) \sin \qty[ \frac{2}{3} ( 3 + \sqrt{3} ) \xi ]
\right) 
\qquad (\xi \gtrsim 1 \ \text{for} \ x \text{-space top-hat}) \,\, .
\end{align}
On the other hand, when $\xi \lesssim \alpha / s$, $\expval{ \delta_R^2 }$ can be universally approximated as
\begin{align}
\expval{ \delta_R^2 } \simeq \frac{\mathcal{A}}{4}  \left( \frac{4}{9} \right)^2 (s^4 - 1) \xi^4 \qquad (\xi \lesssim \alpha/s \ \text{for all the window functions}) \,\, .
\end{align} 
This is because all the Fourier-transformed window functions reach unity for small enough smoothing scale.  
Hence, the difference between the three lines in $\xi \lesssim \alpha / s$ in Figure~\ref{fig:3_1} comes from the window function dependence of $\delta_{\mathrm{th}}$, shown in table~\ref{tab:3_1}.

\begin{table}[h]
\begin{center}
\begin{tabular}{c|ccc}
& Gaussian & $x$-space top-hat & $k$-space top-hat \\
\hline
$\delta_{\mathrm{th}}$ & 0.18 & 0.51 & 0.59
\end{tabular}
\end{center}
\caption{
The window function dependence of the threshold $\delta_{\mathrm{th}}$ at the horizon re-entry~\cite{20Young}. }
\label{tab:3_1}
\end{table}

% \\\\\ \\\\\ Section 4 \\\\\ \\\\\
\section{Window function dependence of the novel mass function}
\label{sec:window_novel}

In this section, we see how much the extra factor $N_{55} \expval{ \delta_R^2 }$ in the novel mass function deviates from unity for each window function, which indicates the new effect from the novel mass function, as we mentioned in section~\ref{sec:basic_formulas}. 
We also discuss the full exponential factor $N_{55} \delta_{\mathrm{th}}^2$ comparing it with the conventional one, $\delta_\text{th}^2/\expval{\delta_R^2}$. 

The factors appearing in $N_{55}$ other than $\expval{ \delta_R^2 }$ are given by 
\begin{enumerate}
\item[(i)]
Gaussian window function:
\begin{align}
\expval{ (\delta_{R, R})^2 }
&= \mathcal{A} \qty( \frac{4}{9} )^2 \frac{1}{R^2} \int^{s \xi }_{\xi} \text{d} u \, u \left( \dv{u} \qty[ u^2 e^{- u^2/4} ] \right)^2 
\nonumber \\
&= - 2 \mathcal{A} \qty( \frac{4}{9} )^2 \frac{1}{R^2} \qty[ \Delta \Gamma_4 - 4 ( \Delta \Gamma_3 - \Delta \Gamma_2 ) ] 
\,\, , 
\label{eq:4_3}
\\
\expval{ \delta_R \delta_{R, R} }
&= \mathcal{A} \qty( \frac{4}{9} )^2 \frac{1}{R} \int^{s \xi }_{\xi} \text{d} u \, u^2 e^{- u^2/4} \, \dv{u} \qty[ u^2 e^{- u^2/4} ]
\nonumber \\
&= 2 \mathcal{A} \qty( \frac{4}{9} )^2 \frac{1}{R} \qty( \Delta \Gamma_3 - 2 \Delta \Gamma_2 )
\,\, ,
\label{eq:4_4}
\end{align}

\item[(ii)]
$x$-space top-hat window function:
\end{enumerate}
\begin{align}
\expval{ ( \delta_{R, R} )^2 }
& = \mathcal{A} \qty( \frac{4}{9} )^2 \frac{1}{R^2} \int^{s \xi }_{\xi} \text{d} u \, u \left( \dv{u} \left[ u^2 \cdot 3 \frac{ \sin u - u \cos u }{ u^3 } \right. \right. \notag \\ 
&\qquad \qquad \qquad \quad \ \, \ \left. \left. \times 3 \frac{ \sin (u / \sqrt{3}) - (u / \sqrt{3}) \cos ( u / \sqrt{3} ) }{ (u / \sqrt{3} )^3 } \right] \right)^2 
\,\, ,
\label{eq:4_7}
\\
\expval{ \delta_R \delta_{R, R} }
&=  \mathcal{A} \qty( \frac{4}{9} )^2 \, \frac{1}{R} \, \int^{s \xi }_{\xi} \text{d} u \, u^2 \cdot 3 \frac{ \sin u - u \cos u }{ u^3 } \cdot 3 \frac{ \sin (u / \sqrt{3}) - (u / \sqrt{3}) \cos ( u / \sqrt{3} ) }{ (u / \sqrt{3} )^3 } \nonumber \\
&\qquad \qquad \quad \ \ \ \, 
\times \dv{u} \left[ u^2 \cdot 3 \frac{ \sin u - u \cos u }{ u^3 } \cdot 3  \frac{ \sin (u / \sqrt{3}) - (u / \sqrt{3}) \cos ( u / \sqrt{3} ) }{ (u / \sqrt{3} )^3 } \right] 
\,\, ,
\label{eq:4_8}
\end{align}

\begin{enumerate}
\item[(iii)]
$k$-space top-hat window function: 
\begin{align}
\expval{ ( \delta_{R, R} )^2 }
&= \mathcal{A} \qty( \frac{4}{9} )^2 \frac{1}{R^2} \int^{s \xi }_{\xi} \text{d} u \, u \left( \dv{u} \qty[ u^2 \Theta(\alpha - u) ] \right)^2 \nonumber \\
&=
\begin{cases}
 \displaystyle \mathcal{A} \qty( \frac{4}{9} )^2 \frac{1}{R^2} ( s^4 -1) \xi^4 
 &\quad (0 < \xi < \alpha/s) \\[10pt]
 \displaystyle \mathcal{A} \qty( \frac{4}{9} )^2 \frac{1}{R^2} \qty[ \alpha^5 \delta_\text{D} (0) + (\alpha^4 - \xi^4) - 4 \alpha^4 \Theta(0) ] 
 &\quad (\alpha / s \leq \xi \leq \alpha) \\
 0 &\quad (\xi > \alpha)
 \end{cases}
\,\, , 
\label{eq:4_5}
\\
\expval{ \delta_R \delta_{R, R} }
&= \mathcal{A} \qty( \frac{4}{9} )^2 \frac{1}{R} \int^{s \xi }_{\xi} \text{d} u \, u^2 \Theta(\alpha - u) \dv{u} \qty[ u^2 \Theta(\alpha - u) ] \nonumber \\
&= 
\begin{cases}
\displaystyle \frac{\mathcal{A}}{2} \qty( \frac{4}{9} )^2 \frac{1}{R} (s^4 -1) \xi^4 
&\quad (0 < \xi < \alpha/s) \\[10pt]
\displaystyle  \frac{\mathcal{A}}{2} \qty( \frac{4}{9} )^2 \frac{1}{R} \left[ \qty( \alpha^4 - \xi^4 ) - 2 \alpha^4 \Theta(0) \right]
&\quad (\alpha / s \leq \xi \leq \alpha) \\
0 &\quad (\xi > \alpha)
\end{cases}
\,\, . 
\label{eq:4_6}
\end{align}

\end{enumerate}
Note that, for $\xi < \alpha / s$, the $k$-space top-hat window function yields $\expval{ \delta_R^2 } \expval{ ( \delta_{R, R} )^2 } - \expval{ \delta_R \delta_{R, R} }^2 = 0$ and hence we cannot define $N_{55}$. 
In this regime, $\delta_{R,R} = 2 \delta_R / R$ holds for the $k$-space top-hat window function, which means $\delta_R > \delta_\text{th}$ and $\delta_{R,R} = 0$ cannot be satisfied at the same time.
Therefore, we can conclude that PBHs are not produced in $\xi < \alpha / s$ for the $k$-space top-hat window function.
The similar issue, the impossibility of defining $N_{55}$, also arises for $\xi > \alpha$, but the fact that $\expval{ \delta_R^2 } = 0$ implies that, again, no PBH formation occurs in the regime.\footnote{
Note that the behaviors of the exponential factor in $\xi < \alpha/s$ or $\xi > \alpha$ for the $k$-space top-hat window function are the consequence of the exact top-hat power spectrum, defined in eq.~(\ref{eq:2_3}).
If we assume the additional power spectrum that is small but scale invariant as $\mathcal P_{\mathcal R} = \mathcal{A} \Theta (k - k_l) \Theta ( - k + k_s )  + \mathcal B$ ($\mathcal B \ll \mathcal A$), $\expval{(\delta_{R,R})^2}$ diverges for all $\xi$, which leads to the same exponential factor as the conventional one regardless of $\xi$.
}
Therefore, for the $k$-space top-hat window function, PBH production is allowed only in $\alpha / s < \xi < \alpha$ and, in this region, the divergence of $\expval{ ( \delta_{R, R} )^2 }$ leads to $N_{55} \expval{ \delta_R^2 } = 1$ and $N_{55} \delta_{\mathrm{th}}^2 = \delta_{\mathrm{th}}^2 / \expval{ \delta_R^2 }$, giving the same exponential factor as the one for the conventional mass function.

\begin{figure}[!h]
\begin{center}
\includegraphics[keepaspectratio, scale=0.7]{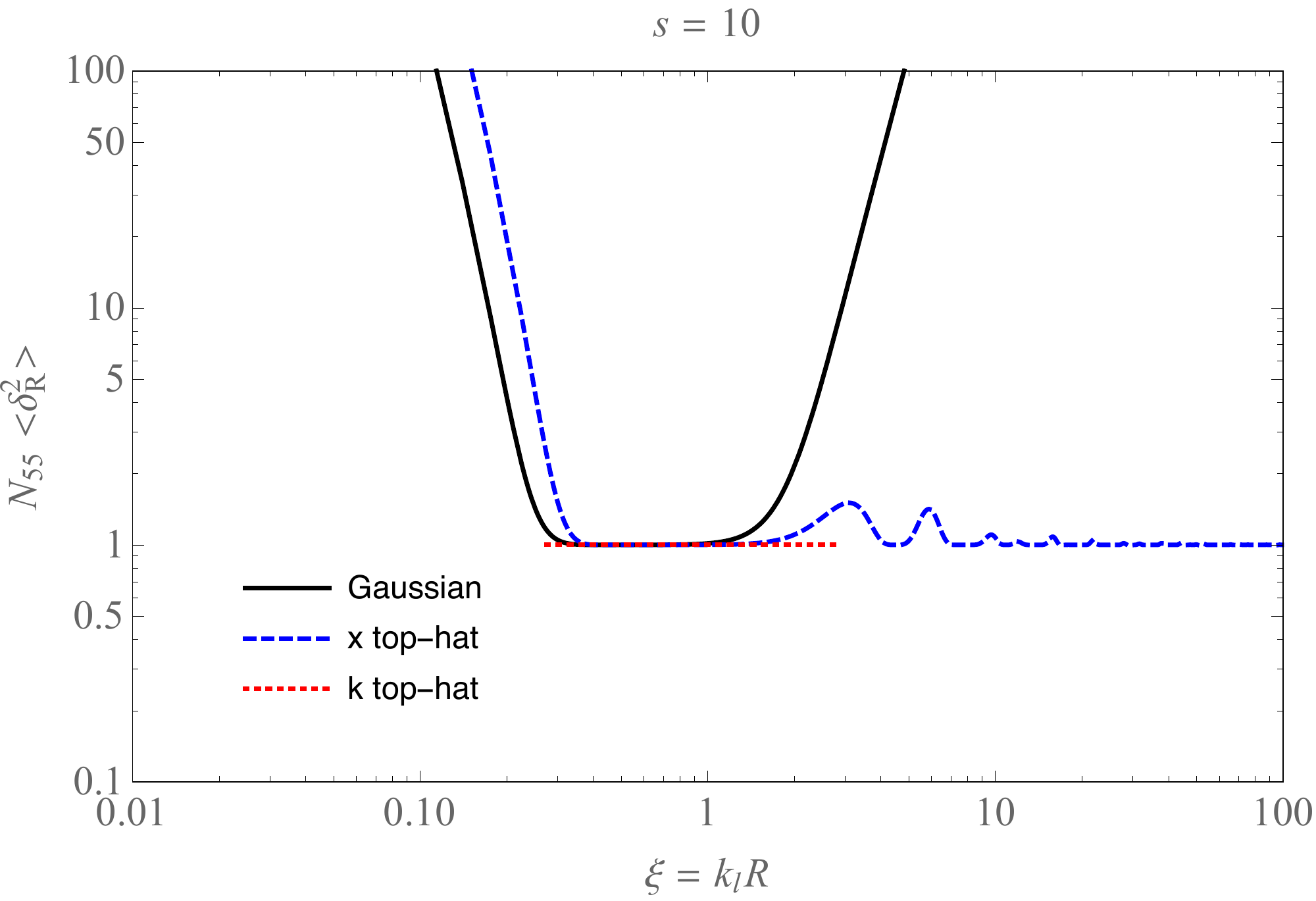}%{20200509_CorrectionFactor_r10.pdf}
\caption{
The behavior of the extra factor in the exponential factor of the novel mass function, given in eq.~(\ref{eq:4_1}), 
with respect to $\xi = k_l R$ for $s =10$.
}
\label{fig:4_1}
\end{center}
\end{figure}

Figure \ref{fig:4_1} shows the extra factor in the exponent of the novel mass function, given by 
\begin{equation}
N_{55} \expval{ \delta_R^2 }
= \frac{ \expval{ \delta_R^2 } \expval{ ( \delta_{R, R} )^2 } }{ \expval{ \delta_R^2 } \expval{ ( \delta_{R, R} )^2 } - \expval{ \delta_R \delta_{R, R} }^2 } \,\, . 
\label{eq:4_1}
\end{equation}
As mentioned above, the result from the $k$-space top-hat window function agrees to the conventional result due to the divergence of $\expval{ ( \delta_{R, R} )^2 }$. 
The other two window functions also give the same magnitudes as the conventional ones, $N_{55} \expval{ \delta_R^2 } \simeq 1$, in $\alpha / s \lesssim \xi \lesssim 1$ due to the fact that $\expval{ \delta_R^2 } \expval{ (\delta_{R, R})^2 } \gg \expval{ \delta_R \delta_{R, R} }^2$ in the regime, which means that $\delta_R$ and $\delta_{R, R}$ are hardly correlated. 
On the other hand, the factor from the Gaussian window function is much larger than unity in $\xi \lesssim \alpha / s$ or $\xi \gtrsim 1$, indicating less PBH formation.
For the $x$-space top-hat window function, while the extra factor behaves similarly to the Gaussian window case in $\xi \lesssim \alpha / s$, it fluctuates a little in $\xi \sim \mathcal O(1)$ and reaches unity again in $\xi \gg 1$.

\begin{figure}[!h]
\begin{center}
\includegraphics[keepaspectratio, scale=0.7]{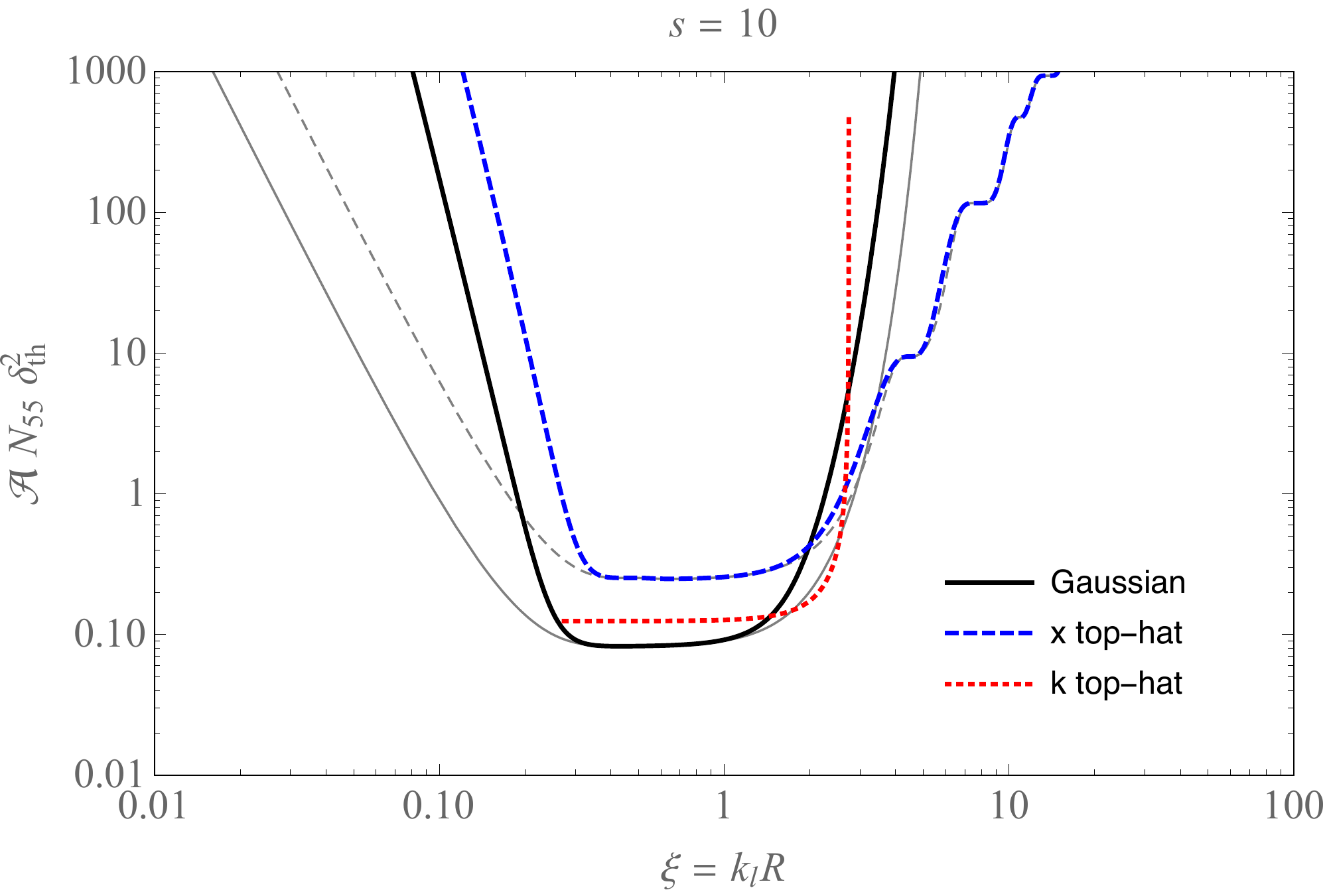}
\caption{
Window function dependence of the exponential factor 
of the novel mass function, given in eq.~(\ref{eq:4_2}), 
with respect to $\xi = k_l R$ for $s =10$. 
For comparison, we also show the conventional factors $\mathcal A\, \delta_{\mathrm{th}}^2 / \expval{ \delta_R^2 }$ from the Gaussian window function (thin gray solid) and the $x$-space top-hat window function (thin gray dashed), which are the same plots as in figure~\ref{fig:3_1}.
} 
\label{fig:4_2}
\end{center}
\end{figure}

Figure \ref{fig:4_2} shows the full exponential factor in the novel mass function, given by 
\begin{equation}
N_{55} \delta_{\mathrm{th}}^2
= \frac{ \expval{ ( \delta_{R, R} )^2 } }{ \expval{ \delta_R^2 } \expval{ ( \delta_{R, R} )^2 } - \expval{ \delta_R \delta_{R, R} }^2 } \delta_{\mathrm{th}}^2 \,\, .
\label{eq:4_2}
\end{equation}
From this figure, we can see that, for $\alpha / s \lesssim \xi \lesssim 1$, the novel mass function predicts the same outcome as the conventional one for all the window functions, as long as we focus on the exponential factor. 
This is true even for the result with the $x$-space top-hat window function in $\xi \gtrsim 1$. 
On the other hand, the results with the Gaussian window function in $\xi \lesssim \alpha / s$ or $\xi \gtrsim 1$ and the $x$-space top-hat window function in $\xi \lesssim \alpha / s$ predict the larger exponent compared to the corresponding factors in the conventional formulation, which means the suppression of the PBH abundance in the regimes and leads to narrow PBH mass spectra.

% \\\\\ \\\\\ Section 5 \\\\\ \\\\\
\section{Conclusion and discussion}
\label{sec:conclusion}

The relation between the power spectrum of the curvature perturbations and the mass function of PBHs has been studied by many authors.
Also, the window function dependence of the mass function has been discussed in refs.~\cite{18AIK,20Young}.
Recently, the novel mass function has been proposed in ref.~\cite{20SY} with the additional criterion that PBH formation should be related to the smoothing scale at which the density contrast becomes maximum.
This novel formulation is free from the questionable assumption in the conventional mass formulation and therefore is superior to the conventional one in some sense.
However, the novel mass function still has the uncertainties originating from the choice of the window function.

In this paper, we have revisited the window function dependence of the PBH mass function, especially focusing on the exponential factor of the mass function, which is a decisive quantity to determine the abundance of PBHs.
We have used the smoothed density contrast $\delta_R$ to describe PBH formation with the top-hat power spectrum $\mathcal{P}_{\mathcal{R}} = \mathcal{A} \Theta (k - k_l) \Theta ( - k + k_s ) $ and considered the three kinds of window functions: the Gaussian, the $x$-space top-hat, and the $k$-space top-hat window functions.
As a result, we have found that all the window functions reproduce the exponential factors of the conventional mass function around the top-hat region ($\alpha / s \lesssim \xi \lesssim 1$).
On the other hand, the Gaussian window function gives the larger exponent than the conventional one in $\xi \lesssim \alpha / s$ or $\xi \gtrsim 1$, and also the x-space top-hat window function does in $\xi \lesssim \alpha / s$. 
This means that the novel mass function predicts the narrower PBH mass spectrum compared to the conventional one, which is consistent with the result in ref.~\cite{Germani:2019zez} which is based on $x$-space top-hat smoothing of other control variables than ours.

\acknowledgments

We thank Teruaki Suyama and Shuiciro Yokoyama for useful communications.
This work was partially supported by JSPS KAKENHI Grant Numbers, 15H02082 and 20H05248.

\appendix

\section{The novel mass function in the 4-dimensional spacetime}
\label{app:4_dim_formula}

%%%%

In this section, we derive the expression of $I(M)$ in the full three-dimensional space with additional variable $R$, extending
the one-dimensional analysis of \cite{20SY}. 
Similar four-dimensional analysis with different variables has been done in \cite{Germani:2019zez}
As we mentioned in eq.~(\ref{eq:2_8}), $I(M)$ in the 4-dimensional ``spacetime'' is defined as 
\begin{equation}
I (M) 
\equiv 
\expval{
\abs{\det H } \Theta (\delta_R - \delta_{\mathrm{th}}) 
\prod_{\alpha = 1}^4 \delta_{\mathrm{D}} (\delta_{R, \alpha}) \Theta (- \lambda_{\alpha}) 
} 
\,\, ,
\label{eq:im_def_app}
\end{equation}
where the Hesse matrix $H$ can be explicitly written as %  $J$ is given as  
\begin{align}
	H = \left(
    \begin{array}{cccc}
      \delta_{R,xx} & \delta_{R,xy} & \delta_{R,xz} & \delta_{R,xR} \\
      \delta_{R,yx} & \delta_{R,yy} & \delta_{R,yz} & \delta_{R,yR} \\        
      \delta_{R,zx} & \delta_{R,zy} & \delta_{R,zz} & \delta_{R,zR} \\  
      \delta_{R,Rx} & \delta_{R,Ry} & \delta_{R,Rz} & \delta_{R,RR}
    \end{array}
  \right).
\end{align}
Here, we transform the delta function and the Heaviside step function as 
\begin{align}
	\delta_\text{D}(\delta_{R,x}) &= \int \frac{\dd \eta_1}{2 \pi} \, e^{i \eta_1 \delta_{R,x}} \,\, , \notag \\
	\delta_\text{D}(\delta_{R,y}) &= \int \frac{\dd \eta_2}{2 \pi} \, e^{i \eta_2 \delta_{R,y}} \,\, , \notag \\
	\delta_\text{D}(\delta_{R,z}) &= \int \frac{\dd \eta_3}{2 \pi} \, e^{i \eta_3 \delta_{R,z}} \,\, , \\
	\delta_\text{D}(\delta_{R,R}) &= \int \frac{\dd \eta_4}{2 \pi} \, e^{i \eta_4 \delta_{R,R}} \,\, , \notag \\
	\Theta(\delta_R - \delta_\text{th}) &= \int^\infty_{\delta_\text{th}} \dd \theta \int \frac{\dd \eta_5}{2\pi} \, e^{i \eta_5 (\delta_R -\theta)} \,\, . \notag
\end{align}
In addition, as done in ref.~\cite{20SY}, we assume $\Theta(-\lambda_\alpha) = 1$ because PBHs are produced by high peaks ($\delta_\text{th} \gg \delta_R$).
Then, we can rewrite eq.~(\ref{eq:im_def_app}) as 
\begin{align}
	I(M) = \qty[ \prod^5_{i = 1} \int \frac{\dd \eta_i}{2 \pi} ] \int^\infty_{\delta_\text{th}} \dd \theta \, e^{-i \eta_5 \theta} 
	\expval{\abs{\det H } e^{i (\eta_1 \delta_{R,x} + \eta_2 \delta_{R,y} + \eta_3 \delta_{R,z} + \eta_4 \delta_{R,R} + \eta_5 \delta_{R}) } }.
	\label{eq:im_formula2}
\end{align}
To perform the integral, we use the following relation: ($x_i$ follow the Gaussian distribution with zero mean and their covariances are given as $\expval{x_i x_j} = P_{ij}$)
\begin{align}
	\expval{x_i x_j x_k x_l e^{i u_m x_m}} =& 
	\expval{x_i x_j x_k x_l \left[ 1 - \frac{1}{2!} (u_k x_k)^2 + \frac{1}{4!} (u_k x_k)^4  - \frac{1}{6!} (u_k x_k)^6 + \cdots \right] }
	\nonumber\\
	=
	&\left[
	P_{ij} P_{kl} + P_{ik} P_{jl} + P_{il} P_{jk} \right. \nonumber \\
	&- \left( P_{ij} P_{km} u_m P_{ln} u_n + P_{ik} P_{jm} u_m P_{ln} u_n + P_{il} P_{jm} u_m P_{kn} u_n \right. \nonumber \\
	&\quad \ \left. + P_{jk} P_{im} u_m P_{ln} u_n + P_{jl} P_{im} u_m P_{kn} u_n + P_{kl} P_{im} u_m P_{jn} u_n  \right) \nonumber \\
	& \left. +  P_{im} u_m P_{jn} u_n  P_{ks} u_s P_{lt} u_t \right] \exp\left(- \frac{1}{2} P_{mn} u_m u_n  \right), 
	\label{eq:ensemble_formula}
\end{align}
where we have killed the irrelevant terms, which are composed of odd-numbered $x_i$, in the first equality because such terms become zero under the Gaussian distribution.
Using this relation, we can rewrite eq.~(\ref{eq:im_formula2}) as 
\begin{align}
	I(M) &= \int^\infty_{\delta_\text{th}} \dd \theta \qty[ \prod^5_{i=1} \int \frac{\dd \eta_i}{2 \pi} ] \left(A + B_{ij} \eta_{i} \eta_{j} + C_{ijkl} \eta_i \eta_j \eta_k \eta_l \right) \exp\left(- \frac{1}{2} L_{ij} \eta_i \eta_j + i v_i \eta_i \right) \notag \\
	& = \frac{1}{\sqrt{\det (2 \pi L)}} 
	\int^\infty_{\delta_\text{th}} \dd \theta \left[ A + B_{ij} N_{ij} - B_{ij} N_{ik} N_{jl} v_k v_l \right. \nonumber \\ 
	& \qquad \qquad \qquad \qquad \qquad \qquad	
	+ 3 C_{ijkl} N_{ij} N_{kl} -6 C_{ijkl} N_{ij} N_{km} v_m N_{lm} v_m \nonumber \\
	& \qquad \qquad \qquad \qquad \qquad \qquad
	\left. + C_{ijkl} N_{im} v_m N_{jm} v_m N_{km} v_m N_{lm} v_m \right] 
	\exp\left(- \frac{1}{2} N_{ij} v_i v_j \right) \nonumber \\
&= \frac{1}{\sqrt{\det (2 \pi L)}} 
\sqrt{ \frac{ 2 }{N_{55} } } 
\notag \\
&\quad 
\times 
\left[
\qty( A + B_{ij} N_{ij} + 3 C_{ijkl} N_{ij} N_{kl} - \frac{(B_{ij} + 6 C_{ijkl} N_{kl} ) N_{i5} N_{j5} }{N_{55}} 
+ \frac{ 3 C_{ijkl} N_{i5} N_{j5} N_{k5} N_{l5} }{N_{55}^2} 
)
\right.
\notag \\
&\qquad \quad 
\quad 
\times 
\frac{\sqrt{\pi}}{2} \mathrm{erfc} \qty( \sqrt{ \frac{N_{55}}{2} } \, \delta_{\mathrm{th}} ) 
\notag \\ 
&\qquad \quad 
+ \frac{\delta_{\mathrm{th}}}{\sqrt{2 N_{55}}} 
\qty( 
( B_{ij} + 6 C_{ijkl} N_{kl} ) N_{i5} N_{j5} 
+ \frac{ \delta_{\mathrm{th}} (3 + N_{55} \delta_{\mathrm{th}}^2 ) C_{ijkl} N_{i5} N_{j5} N_{k5} N_{l5} }{N_{55}} ) 
\notag \\
&\qquad \quad 
\quad 
\times 
\left.
\exp \qty( - \frac{N_{55}}{2} \delta_{\mathrm{th}}^2 ) 
\right] \,\, ,
\label{eq:im_final}
\end{align}
where $L_{ij}$ is defined in eq.~(\ref{eq:2_10}),
$v_i$ is given as $v_i = (0,0,0,0,-\theta)$ and $N_{ij}$ is the inverse matrix of $L_{ij}$, given as 
%%

\begin{comment}
\begin{align}
	N_{ij} &= \left(
    \begin{array}{ccccc}
      \displaystyle \frac{1}{\expval{(\delta_{R,x})^2}} & 0 & 0 & 0 & 0 \\
      0 & \displaystyle \frac{1}{\expval{(\delta_{R,y})^2}} & 0 & 0 & 0 \\        
      0 & 0 & \displaystyle \frac{1}{\expval{(\delta_{R,z})^2}} & 0 & 0 \\              
      0 & 0 & 0 & \displaystyle \frac{\expval{\delta_{R}^2}}{\expval{\delta_{R}^2}\expval{(\delta_{R,R})^2} - \expval{\delta_{R,R} \delta_R}^2 } & \displaystyle  -\frac{\expval{\delta_{R,R} \delta_R}}{\expval{\delta_{R}^2}\expval{(\delta_{R,R})^2} - \expval{\delta_{R,R} \delta_R}^2 } \\       
      0 & 0 & 0 & \displaystyle -\frac{\expval{\delta_{R,R} \delta_R}}{\expval{\delta_{R}^2}\expval{(\delta_{R,R})^2} - \expval{\delta_{R,R} \delta_R}^2 } & \displaystyle \frac{\expval{(\delta_{R,R})^2}}{\expval{\delta_{R}^2}\expval{(\delta_{R,R})^2} - \expval{\delta_{R,R} \delta_R}^2 }  \\             
    \end{array}
  \right).
\end{align}
\end{comment}

\begin{equation}
N = \mqty( N_{11} & & & & \\
& N_{22} & & & \\
& & N_{33} & & \\
& & & N_{44} & N_{45} \\
& & & N_{54} & N_{55} ) \,\, ,
\end{equation}
where 
\begin{align}
\mqty( N_{11} & & \\ & N_{22} & \\ & & N_{33} )
&= \mqty( 
\displaystyle \frac{1}{\expval{(\delta_{R,x})^2}} & & \\
      & \displaystyle \frac{1}{\expval{(\delta_{R,y})^2}} &  \\        
      & & \displaystyle \frac{1}{\expval{(\delta_{R,z})^2}} ) \,\, , \\
\mqty( N_{44} & N_{45} \\ N_{54} & N_{55} ) 
&= \mqty( 
\displaystyle \ \ \frac{\expval{\delta_{R}^2}}{\expval{\delta_{R}^2}\expval{(\delta_{R,R})^2} - \expval{\delta_{R,R} \delta_R}^2 } 
& 
\displaystyle \ \ -\frac{\expval{\delta_{R,R} \delta_R}}{\expval{\delta_{R}^2}\expval{(\delta_{R,R})^2} - \expval{\delta_{R,R} \delta_R}^2 } 
\\       
\displaystyle -\frac{\expval{\delta_{R,R} \delta_R}}{\expval{\delta_{R}^2}\expval{(\delta_{R,R})^2} - \expval{\delta_{R,R} \delta_R}^2 } 
& 
\displaystyle \ \ \ \ \frac{\expval{(\delta_{R,R})^2}}{\expval{\delta_{R}^2}\expval{(\delta_{R,R})^2} - \expval{\delta_{R,R} \delta_R}^2 } 
) \,\, .
\end{align}      
$A$ in eq.~(\ref{eq:im_final}) is given as
\begin{equation}
A = 
\qty( \expval{ \delta_{R, RR} \delta_{R, xx} } - \expval{ ( \delta_{R, Rx} )^2 } ) 
\qty( \expval{ \delta_{R, yy} \delta_{R, zz} } - \expval{ ( \delta_{R, yz} )^2 } ) 
+ (x \leftrightarrow y ) 
+ (x \leftrightarrow z ) 
\,\, ,
\label{eq:appx_a2}
\end{equation}
$B_{ij}$ is the $5 \times 5$ symmetric and block diagonal matrix whose nonzero components are given by 
\begin{align}
B_{11} 
&= \expval{ \delta_{R, x} \delta_{R, Rx} }^2 
\qty( \expval{ \delta_{R, yy} \delta_{R, zz} } - \expval{ \delta_{R, yz} }^2 ) \,\, , \\
B_{22} 
&= \expval{ \delta_{R, y} \delta_{R, Ry} }^2 
\qty( \expval{ \delta_{R, zz} \delta_{R, xx} } - \expval{ \delta_{R, zx} }^2 ) \,\, , \\
B_{33} 
&= \expval{ \delta_{R, z} \delta_{R, Rz} }^2 
\qty( \expval{ \delta_{R, xx} \delta_{R, yy} } - \expval{ \delta_{R, xy} }^2 ) \,\, , \\
- B_{44} 
&= \left[ 
\expval{ \delta_{R, R} \delta_{R, RR} } \expval{ \delta_{R, R}  \delta_{R, xx} } 
\qty( 
\expval{ \delta_{R, yy} \delta_{R, zz} } - \expval{ \delta_{R, yz} }^2 
) 
\right. \notag \\
&\quad \, +
\left. 
\expval{ \delta_{R, R} \delta_{R, yy} } \expval{ \delta_{R, R}  \delta_{R, zz} } 
\qty( 
\expval{ \delta_{R, RR} \delta_{R, xx} } - \expval{ \delta_{R, Rx} }^2 
) 
\right] 
\notag \\
&\quad \, + ( x \leftrightarrow y ) + (x \leftrightarrow z ) \,\, , \\
- B_{55} 
&= \left[ 
\expval{ \delta_R \delta_{R, RR} } \expval{ \delta_R \delta_{R, xx} } 
\qty( 
\expval{ \delta_{R, yy} \delta_{R, zz} } - \expval{ \delta_{R, yz} }^2 
) 
\right. \notag \\
&\quad \, +
\left. 
\expval{ \delta_R \delta_{R, yy} } \expval{ \delta_R \delta_{R, zz} } 
\qty( 
\expval{ \delta_{R, RR} \delta_{R, xx} } - \expval{ \delta_{R, Rx} }^2 
) 
\right] 
\notag \\
&\quad \,
+ ( x \leftrightarrow y ) + (x \leftrightarrow z ) \,\, , \\
- 2 B_{45} 
&= \left[ 
\qty( 
\expval{ \delta_R \delta_{R, RR} } \expval{ \delta_{R, R} \delta_{R, xx} } 
+ \expval{ \delta_R \delta_{R, xx} } \expval{ \delta_{R, R} \delta_{R, RR} } 
) \qty( 
\expval{ \delta_{R, yy} \delta_{R, zz} } - \expval{ \delta_{R, yz} }^2 
) 
\right. 
\notag \\
&\quad \, 
\left.
+ 
\qty( 
\expval{ \delta_R \delta_{R, yy} } \expval{ \delta_{R, R} \delta_{R, zz} } 
+ \expval{ \delta_R \delta_{R, zz} } \expval{ \delta_{R, R} \delta_{R, yy} } 
) \qty( 
\expval{ \delta_{R, RR} \delta_{R, xx} } - \expval{ \delta_{R, Rx} }^2 
) 
\right] 
\notag \\
&\quad \, 
+ ( x \leftrightarrow y ) + (x \leftrightarrow z ) \,\, ,
\end{align}
and $C_{ijkl}$ is the tensor invariant under an arbitrary exchange between $i$, $j$, $k$, and $l \in \{ 1, \dots, 5 \}$ whose nonzero components are given by (for $i \leq j \leq k \leq l$)
\begin{align}
- 6 C_{1144} 
&= \expval{ \delta_{R, x} \delta_{R, Rx} }^2 
\expval{ \delta_{R, R} \delta_{R, yy} } 
\expval{ \delta_{R, R} \delta_{R, zz} } \,\, ,
\\
- 6 C_{2244} 
&= \expval{ \delta_{R, y} \delta_{R, Ry} }^2 
\expval{ \delta_{R, R} \delta_{R, zz} } 
\expval{ \delta_{R, R} \delta_{R, xx} } \,\, ,
\\
- 6 C_{3344} 
&= \expval{ \delta_{R, z} \delta_{R, Rz} }^2 
\expval{ \delta_{R, R} \delta_{R, xx} } 
\expval{ \delta_{R, R} \delta_{R, yy} } \,\, , 
\\
- 6 C_{1155} 
&= \expval{ \delta_{R, x} \delta_{R, Rx} }^2 
\expval{ \delta_{R} \delta_{R, yy} } 
\expval{ \delta_{R} \delta_{R, zz} } \,\, ,
\\
- 6 C_{2244} 
&= \expval{ \delta_{R, y} \delta_{R, Ry} }^2 
\expval{ \delta_{R} \delta_{R, zz} } 
\expval{ \delta_{R} \delta_{R, xx} } \,\, ,
\\
- 6 C_{3344} 
&= \expval{ \delta_{R, z} \delta_{R, Rz} }^2 
\expval{ \delta_{R} \delta_{R, xx} } 
\expval{ \delta_{R} \delta_{R, yy} } \,\, , 
\\
-12 C_{1145} 
&= \expval{ \delta_{R, x} \delta_{R, Rx} }^2 
\qty( 
\expval{ \delta_R \delta_{R, yy} } \expval{ \delta_{R, R} \delta_{R, zz} } 
+ \expval{ \delta_R \delta_{R, zz} } \expval{ \delta_{R, R} \delta_{R, yy} } 
) \,\, , \\
-12 C_{2245} 
&= \expval{ \delta_{R, y} \delta_{R, Ry} }^2 
\qty( 
\expval{ \delta_R \delta_{R, zz} } \expval{ \delta_{R, R} \delta_{R, xx} } 
+ \expval{ \delta_R \delta_{R, xx} } \expval{ \delta_{R, R} \delta_{R, zz} } 
) \,\, , \\
-12 C_{1145} 
&= \expval{ \delta_{R, z} \delta_{R, Rz} }^2 
\qty( 
\expval{ \delta_R \delta_{R, xx} } \expval{ \delta_{R, R} \delta_{R, yy} } 
+ \expval{ \delta_R \delta_{R, yy} } \expval{ \delta_{R, R} \delta_{R, xx} } 
) \,\, , \\ 
C_{4444} 
&= \expval{ \delta_{R, R} \delta_{R, RR} } 
\expval{ \delta_{R, R} \delta_{R, xx} } 
\expval{ \delta_{R, R} \delta_{R, yy} } 
\expval{ \delta_{R, R} \delta_{R, zz} } \,\, , \\
C_{5555} 
&= \expval{ \delta_{R} \delta_{R, RR} } 
\expval{ \delta_{R} \delta_{R, xx} } 
\expval{ \delta_{R} \delta_{R, yy} } 
\expval{ \delta_{R} \delta_{R, zz} } \,\, , \\
6 C_{4455} 
&= 
\expval{ \delta_R \delta_{R, RR} } 
\qty[ 
\expval{ \delta_R \delta_{R, xx} } \expval{ \delta_{R, R} \delta_{R, yy} } \expval{ \delta_{R, R} \delta_{R, zz} } 
+ ( x \leftrightarrow y ) + ( x \leftrightarrow z ) 
] 
\notag \\
&\quad \, + 
\expval{ \delta_{R, R} \delta_{R, RR} } 
\qty[ 
\expval{ \delta_{R, R} \delta_{R, xx} } \expval{ \delta_R \delta_{R, yy} } \expval{ \delta_R \delta_{R, zz} } 
+ ( x \leftrightarrow y ) + ( x \leftrightarrow z ) 
] 
\,\, , \\
4 C_{4445} 
&= \expval{ \delta_R \delta_{R, RR} } \expval{ \delta_{R, R} \delta_{R, xx} } 
\expval{ \delta_{R, R} \delta_{R, yy} } \expval{ \delta_{R, R} \delta_{R, zz} } \notag \\
&\quad \, 
+ \expval{ \delta_{R, R} \delta_{R, RR} } 
\qty[ 
\expval{ \delta_R \delta_{R, xx} } 
\expval{ \delta_{R, R} \delta_{R, yy} } \expval{ \delta_{R, R} \delta_{R, zz} } 
+ ( x \leftrightarrow y ) + ( x \leftrightarrow z ) 
] \,\, , \\
4 C_{4555} 
&= \expval{ \delta_{R, R} \delta_{R, RR} } \expval{ \delta_{R} \delta_{R, xx} } 
\expval{ \delta_{R} \delta_{R, yy} } \expval{ \delta_{R} \delta_{R, zz} } \notag \\
&\quad \, + 
\expval{ \delta_{R} \delta_{R, RR} } 
\qty[ 
\expval{ \delta_{R, R} \delta_{R, xx} } 
\expval{ \delta_{R} \delta_{R, yy} } \expval{ \delta_{R} \delta_{R, zz} } 
+ ( x \leftrightarrow y ) + ( x \leftrightarrow z ) 
] \,\, . 
\end{align}

\bibliography{references}
\bibliographystyle{junsrt}

\begin{comment}

\end{comment}

\end{document}